\documentclass[aps,preprint]{revtex4}
\usepackage{amssymb,epsf}
\usepackage{amsmath,amsfonts}
\usepackage{graphicx}
\usepackage{epsfig}
\usepackage{epstopdf}
\usepackage{graphicx}

\begin{document}

\author{N. Farhangkhah $^{1}$and M. H. Dehghani $^{2,3}$\thanks{%
mhd@shirazu.ac.ir}}
\title{Lovelock black holes with nonmaximally symmetric horizons}
\affiliation{$^1$ Department of Physics, Shiraz Branch, Islamic Azad University, Shiraz 71993,
Iran\\
$^2$Research Institute for Astrophysics and Astronomy of Maragha (RIAAM),
Maragha, Iran\\
$^3$Physics Department and Biruni Observatory, College of Sciences, Shiraz
University, Shiraz 71454, Iran}

\begin{abstract}
We present a new class of black hole solutions in third-order Lovelock
gravity whose horizons are Einstein space with two supplementary conditions
on their Weyl tensors. These solutions are obtained with the advantage of
higher curvature terms appearing in Lovelock gravity. We find that while the
solution of third-order Lovelock gravity with constant-curvature horizon in
the absence of a mass parameter is the anti de Sitter (AdS) metric, this kind
of solution with nonconstant-curvature horizon is only asymptotically AdS
and may have horizon. We also find that one may have an extreme black hole
with non-constant curvature horizon whose Ricci scalar is zero or a positive
constant, while there is no such black hole with constant-curvature
horizon. Furthermore, the thermodynamics of the black holes in the two cases
of constant- and nonconstant-curvature horizons are different drastically.
Specially, we consider the thermodynamics of black holes with vanishing
Ricci scalar and find that in contrast to the case of black holes of
Lovelock gravity with constant-curvature horizon, the area law of entropy is
not satisfied. Finally, we investigate the stability of these black holes
both locally and globally and find that while the black holes with constant
curvature horizons are stable both locally and globally, those with
nonconstant-curvature horizons have unstable phases.
\end{abstract}

\maketitle

\section{Introduction}

Higher-order curvature theories of gravity have gained a lot of attention.
Not surprisingly, the mere extension of general relativity in higher
dimensions can immediately lead to a wide variety of alternative theories of
gravity whose actions contain higher-order curvature terms. The inclusion of
higher curvature terms in the gravitational action increases further the
diversity of the models available and gives rise to a rich phenomenology,
which is actively investigated these days. Amongst the higher curvature
theories of gravity, Lovelock theory \cite{Lovelock} which is the most
general second-order gravity theory in higher-dimensional spacetimes has
attracted a lot of attention. The action imposed in this theory is
consistent with the corrections inspired by string theory to
Einstein-Hilbert action \cite{string}. The most extensively researches are
done on Einstein-Gauss-Bonnet gravity with second-order curvature
corrections \cite{GB1, GB2, GB3}. Although the Lagrangian and field
equations look complicated in third-order Lovelock gravity, there are a
large number of works on introducing and discussing various exact black hole
solutions of third-order Lovelock gravity \cite{Lovelockex1, Lovelockex2,
Lovelockex3}. It is known that the second-order Lovelock gravity admits
supersymmetric extension \cite{SupGB}, while all the higher orders of this
theory have only the necessary condition of supersymmetric extension \cite%
{SupLg}. Throughout the recent years, most of the interesting holographic
aspects of Lovelock gravity have been studied \cite{HolLg3}. Recently, some
works have been extended to general Lovelock gravity to investigate the
solutions and their properties \cite{Lovelockg1, Lovelockg2, Lovelockg3}.

Although most of the known black hole solutions of Lovelock gravity are
those with curvature constant horizons, one may raise the question of having
black hole solutions with nonconstant curvature. Here, specially, we
investigate black hole solutions with Einstein horizon. In four dimensions,
the first explicit inhomogeneous compact Einstein metric was constructed by
Page \cite{Page} and a higher-dimensional version of the method of Page was
given in \cite{Hashimoto}. Bohm constructed an infinite family of
inhomogeneous metrics with positive scalar curvature on products of spheres
\cite{Bohm}. After that, examples in higher-dimensional spacetimes have been
worked in Refs. \cite{Gibbons1, LuPa, Gauntlett}. The properties of such
Einstein manifolds are investigated in five and higher dimensions in Refs.
\cite{Gibbons2} and \cite{Gibbons3, Gibbons4}, respectively.

In this paper we are interested in black hole solutions of third-order
Lovelock gravity whose horizons are Einstein manifolds of nonconstant
curvature. In Einstein gravity, no new solution can be obtained with
nonconstant-curvature boundary. This is due to the fact that the Einstein
equation deals with the Ricci tensor and therefore the Weyl tensor does not
appear in the field equation. On the other hand, the Riemann tensor has direct
contribution in the field equation of Lovelock gravity, and therefore the Weyl
tensor appears in the field equation of Lovelock gravity. In Ref. \cite%
{Dotti} Dotti and Gleiser obtained a condition on an invariant built out of
the Weyl tensor in Gauss-Bonnet gravity when the horizon is an Einstein
manifold. This constraint appears in the metric and consequently changes the
properties of the spacetime. The properties of such static and dynamical
solutions in Einstein-Gauss-Bonnet (EGB) gravity have been investigated in
\cite{Maeda}. Also the magnetic black hole that has space with such specific
condition was obtained in \cite{Maeda2}. While the base manifold of
black hole solutions in EGB gravity with a generic value of the coupling
constant must be necessarily Einstein, the boundary admits a wider class of
geometries in the special case when the coupling constant is such that the
theory admits a unique maximally symmetric solution \cite{Dot2}.
The Birkhoff's theorem in six-dimensional EGB gravity for the case
of nonconstant-curvature horizons with various features has been
investigated in \cite{Bog}. In Ref. \cite{Oliv1}, it is shown that the
horizons of black holes of Lovelock gravity in the Chern-Simons case \cite{LBI}
in odd dimensions are not restricted. Some specific examples of black holes
of Lovelock-Born-Infeld gravity \cite{LBI} with non-Einstein horizons in
even dimensions were found in \cite{Can}. In Ref. \cite{Oliv2}, it is shown
that the base manifolds of these black hole solutions possess more than one
curvature scale provided avoiding tensor restrictions on the base manifold
and allowing at most a reduced set of scalar constraints on it. While all
the Lovelock coefficients in Lovelock-Chern-Simons and Lovelock-Born-Infeld
gravity are given in term of the cosmological constant, here we do not impose
any condition on the coupling constants of Lovelock gravity and generalize
the idea of Ref. \cite{Dotti} to the case of third-order Lovelock gravity
with arbitrary Lovelock coefficients. That is, we like to obtain the black
hole solutions of third-order Lovelock gravity with arbitrary coupling
constants and nonconstant-curvature horizons. We predict that the appearing
higher-curvature terms in third-order Lovelock gravity, even more sharply,
may cause novel changes in the properties of the spacetime. This is the
motivation for obtaining new black hole solutions in third-order Lovelock
gravity with nonconstant-curvature horizons and investigating their thermodynamic
properties.

The paper is organized as follows. In the following section we begin with a
brief review of the field equation in third-order Lovelock gravity and obtain
the equations getting use of the expressions in warped geometry for our
spacetime ansatz. In Sec. \ref{Bla} we obtain the black hole solutions
and discuss their properties. In Sec. \ref{The}, we calculate the
thermodynamic quantities of the solutions and investigate the first law of
thermodynamics. Section \ref{Stab} is devoted to the analysis of local and
global stabilities by considering the variation of temperature versus
entropy and the free energy for the special case of $\kappa=0$. We finish
our paper with some concluding remarks.

\section{Field Equations}

The most fundamental assumption in standard general relativity is the
requirement that the field equations should be generally covariant and
contain at most a second-order derivative of the metric. Based on this
principle, the most general classical theory of gravitation in $n$
dimensions is Lovelock gravity \cite{Lovelock}. The Lovelock equation up to
third-order terms in vacuum may be written as
\begin{equation}
\mathcal{G}_{\mu \nu } \equiv -\Lambda g_{\mu \nu }+G_{\mu \nu
}^{(1)}+\sum_{p=2}^{3}\alpha _{i}\left( H_{\mu \nu }^{(p)}-\frac{1}{2}g_{\mu
\nu }\mathcal{L}^{(p)}\right) =0,  \label{Geq}
\end{equation}
where $\Lambda $ is the cosmological constant, $\alpha _{p}$'s are Lovelock
coefficients,$G_{\mu \nu }^{(1)}$ is just the Einstein tensor, $\mathcal{L}%
^{(2)}=R_{\mu \nu \gamma \delta }R^{\mu \nu \gamma \delta }-4R_{\mu \nu
}R^{\mu \nu }+R^{2} $ is the Gauss-Bonnet Lagrangian,
\begin{eqnarray}
\mathcal{L}^{(3)} &=&2R^{\mu \nu \sigma \kappa }R_{\sigma \kappa \rho \tau
}R_{\phantom{\rho \tau }{\mu \nu }}^{\rho \tau }+8R_{\phantom{\mu
\nu}{\sigma \rho}}^{\mu \nu }R_{\phantom {\sigma \kappa} {\nu \tau}}^{\sigma
\kappa }R_{\phantom{\rho \tau}{ \mu \kappa}}^{\rho \tau }+24R^{\mu \nu
\sigma \kappa }R_{\sigma \kappa \nu \rho }R_{\phantom{\rho}{\mu}}^{\rho }
\notag \\
&&+3RR^{\mu \nu \sigma \kappa }R_{\sigma \kappa \mu \nu }+24R^{\mu \nu
\sigma \kappa }R_{\sigma \mu }R_{\kappa \nu }+16R^{\mu \nu }R_{\nu \sigma
}R_{\phantom{\sigma}{\mu}}^{\sigma }-12RR^{\mu \nu }R_{\mu \nu }+R^{3}
\label{Lag3}
\end{eqnarray}
is the third-order Lovelock Lagrangian, and $H_{\mu \nu }^{(2)}$ and $H_{\mu
\nu }^{(3)}$ are
\begin{equation}
H_{\mu \nu }^{(2)}=2(R_{\mu \sigma \kappa \tau }R_{\nu }^{\phantom{\nu}%
\sigma \kappa \tau }-2R_{\mu \rho \nu \sigma }R^{\rho \sigma }-2R_{\mu
\sigma }R_{\phantom{\sigma}\nu }^{\sigma }+RR_{\mu \nu }),  \label{Love2}
\end{equation}
\begin{eqnarray}
H_{\mu \nu }^{(3)} &=&-3(4R^{\tau \rho \sigma \kappa }R_{\sigma \kappa
\lambda \rho }R_{\phantom{\lambda }{\nu \tau \mu}}^{\lambda }-8R_{%
\phantom{\tau \rho}{\lambda \sigma}}^{\tau \rho }R_{\phantom{\sigma
\kappa}{\tau \mu}}^{\sigma \kappa }R_{\phantom{\lambda }{\nu \rho \kappa}%
}^{\lambda }+2R_{\nu }^{\phantom{\nu}{\tau \sigma \kappa}}R_{\sigma \kappa
\lambda \rho }R_{\phantom{\lambda \rho}{\tau \mu}}^{\lambda \rho }  \notag \\
&&-R^{\tau \rho \sigma \kappa }R_{\sigma \kappa \tau \rho }R_{\nu \mu }+8R_{%
\phantom{\tau}{\nu \sigma \rho}}^{\tau }R_{\phantom{\sigma \kappa}{\tau \mu}%
}^{\sigma \kappa }R_{\phantom{\rho}\kappa }^{\rho }+8R_{\phantom
{\sigma}{\nu \tau \kappa}}^{\sigma }R_{\phantom {\tau \rho}{\sigma \mu}%
}^{\tau \rho }R_{\phantom{\kappa}{\rho}}^{\kappa }  \notag \\
&&+4R_{\nu }^{\phantom{\nu}{\tau \sigma \kappa}}R_{\sigma \kappa \mu \rho
}R_{\phantom{\rho}{\tau}}^{\rho }-4R_{\nu }^{\phantom{\nu}{\tau \sigma
\kappa }}R_{\sigma \kappa \tau \rho }R_{\phantom{\rho}{\mu}}^{\rho
}+4R^{\tau \rho \sigma \kappa }R_{\sigma \kappa \tau \mu }R_{\nu \rho
}+2RR_{\nu }^{\phantom{\nu}{\kappa \tau \rho}}R_{\tau \rho \kappa \mu }
\notag \\
&&+8R_{\phantom{\tau}{\nu \mu \rho }}^{\tau }R_{\phantom{\rho}{\sigma}%
}^{\rho }R_{\phantom{\sigma}{\tau}}^{\sigma }-8R_{\phantom{\sigma}{\nu \tau
\rho }}^{\sigma }R_{\phantom{\tau}{\sigma}}^{\tau }R_{\mu }^{\rho }-8R_{%
\phantom{\tau }{\sigma \mu}}^{\tau \rho }R_{\phantom{\sigma}{\tau }}^{\sigma
}R_{\nu \rho }  \notag \\
&&-4RR_{\phantom{\tau}{\nu \mu \rho }}^{\tau }R_{\phantom{\rho}\tau }^{\rho
}+4R^{\tau \rho }R_{\rho \tau }R_{\nu \mu }-8R_{\phantom{\tau}{\nu}}^{\tau
}R_{\tau \rho }R_{\phantom{\rho}{\mu}}^{\rho }+4RR_{\nu \rho }R_{%
\phantom{\rho}{\mu }}^{\rho }-R^{2}R_{\mu \nu }),  \label{Love3}
\end{eqnarray}
respectively.

We take the $n$-dimensional manifold $\mathcal{M}^{n}$ to be a warped
product of a two-dimensional Riemannian submanifold $\mathcal{M}^{2}$ with
the following line element

\begin{equation}
ds^{2}=-f(r)dt^{2}+g(r)dr^{2}.  \label{metric1}
\end{equation}
and an $(n-2)$-dimensional submanifold $\mathcal{K}^{(n-2)}$ with the metric

\begin{equation}
ds^{2}=r^{2}\gamma _{ij}(z)dz^{i}dz^{j}.  \label{metric2}
\end{equation}%
We assume the submanifold $\mathcal{K}^{(n-2)}$ with the unit metric $\gamma
_{ij}$ to be an Einstein manifold with nonconstant curvature and volume $%
V_{n-2}$, where $i,j=2...n-1$. We use tilde for the tensor components of the
submanifold $\mathcal{K}^{(n-2)}$ through the paper. The Ricci tensor, Ricci
scalar and Einstein tensor of the Einstein manifold $\mathcal{K}^{(n-2)}$ are

\begin{eqnarray}
\tilde{R}{_{ij}} &=&\kappa (n-3)\gamma _{ij},\text{ \ \ \ }\tilde{R}=\kappa
(n-2)(n-3), \\
\widetilde{G}{_{ij}} &{=}&\kappa (n-3)\left( 1-\frac{n-2}{2}\right) \gamma
_{ij},
\end{eqnarray}%
respectively. It is worth mentioning that Einstein metrics are vacuum
solutions of Einstein's theory of gravity only in three and four dimensions. The
Riemann tensor of the Einstein manifolds should satisfy
\begin{equation}
\tilde{R}{_{ij}}^{kl}=\tilde{C}{_{ij}}^{kl}+\kappa ({\delta _{i}}^{k}{\delta
_{j}}^{l}-{\delta _{i}}^{l}{\delta _{j}}^{k})  \label{Riem Ten}
\end{equation}%
with $\kappa $ being the sectional curvature and $\tilde{C}{_{ij}}^{kl}$\ is
the Weyl tensor of $\mathcal{K}^{(n-2)}$. Using the expressions in warped
geometry, the sectional components of the field equation (\ref{Geq}) are
calculated to be
\begin{equation}
\mathcal{G}_{i}^{j}=\frac{2\hat{\alpha}_{2}{\tilde{C}_{ki}}^{nl}{\tilde{C}%
_{nl}}^{kj}}{r^{4}}-\frac{3\hat{\alpha}_{3}(4\tilde{C}^{nmkl}\tilde{C}_{klpm}%
{\tilde{C}^{pj}}_{ni}-8{\tilde{C}^{nm}}_{pk}{\tilde{C}^{kl}}_{ni}{\tilde{C}%
^{pj}}_{ml}+2\tilde{C}^{jnkl}\tilde{C}_{klpm}{\tilde{C}^{pm}}_{ni})}{2r^{6}},%
\text{ \ \ \ }(i\neq j)  \label{Gij}
\end{equation}%
\begin{eqnarray}
\mathcal{G}_{i}^{i} &=&\frac{(n-2)}{4g^{4}f^{2}r^{5}}\{2fgr[r^{4}g^{2}+2\hat{%
\alpha}_{2}r^{2}(kg-1)g+\hat{\alpha}_{3}(kg-1)^{2}]f^{\prime \prime
}-gr[r^{4}g^{2}+2\hat{\alpha}_{2}r^{2}(kg-1)g  \notag \\
&&+\hat{\alpha}_{3}(kg-1)^{2}]f^{\prime }{}^{2}-f[((r^{4}+2k\hat{\alpha}%
_{2}r^{2}+\hat{\alpha}_{3})g^{2}+(-6\hat{\alpha}_{2}r^{2}-6k\hat{\alpha}%
_{3})g+5\hat{\alpha}^{3})rg^{\prime }  \notag \\
&&-g(2(n-3)r^{4}g^{2}+4(n-5)\hat{\alpha}_{2}r^{2}(kg-1)g+2(n-7)\hat{\alpha}%
_{3}(kg-1)^{2})]f^{\prime }-2(n-3)  \notag \\
&&[g^{\prime }(r^{4}g^{2}+2g\hat{\alpha}_{2}\frac{n-5}{n-3}r^{2}(kg-1)+\frac{%
n-7}{n-3}\hat{\alpha}_{3}(kg-1)^{2})+(n-4)g(kg-1)  \notag \\
&&(r^{3}g^{2}+\frac{n-5}{10}\hat{\alpha}_{2}r(kg-1)g+\frac{n-8}{n(n-4)}\hat{%
\alpha}_{3}r^{(8-n)}(kg-1)^{2})]f^{2}\}  \notag \\
&&-\left\{ \frac{(n-1)(n-2)\hat{\alpha}_{0}}{2}+\frac{\hat{\alpha}_{2}{%
\tilde{C}_{km}}^{ln}{\tilde{C}_{ln}}^{km}}{2r^{4}}-\frac{\hat{\alpha}_{3}(%
\tilde{C}^{nqkl}\tilde{C}_{klpm}{\tilde{C}^{pm}}_{nq}+4{\tilde{C}^{nm}}_{pk}{%
\tilde{C}^{kl}}_{nr}{\tilde{C}^{pr}}_{ml})}{r^{6}}\right\}   \notag \\
&&+\frac{2\hat{\alpha}_{2}\sum_{kln}{\tilde{C}_{ki}}^{ln}{\tilde{C}_{ln}}%
^{ki}}{r^{4}}+\frac{8\hat{\alpha}_{3}\sum_{klmnp}(\tilde{C}^{inkl}\tilde{C}%
_{klpm}{\tilde{C}^{pm}}_{in}+4{\tilde{C}^{nm}}_{pk}{\tilde{C}^{kl}}_{ni}{%
\tilde{C}^{pi}}_{ml})}{r^{6}};\text{ no sum on }i,  \notag
\end{eqnarray}%
where $\hat{\alpha}_{p}$ are defined as $\hat{\alpha}_{0}\equiv -2\Lambda
/(n-1)(n-2)$, $\hat{\alpha}_{2}\equiv (n-3)(n-4)\alpha _{2}$ and $\hat{\alpha%
}_{3}\equiv (n-3)!\alpha _{3}/(n-7)!$ for simplicity. In vacuum, $\mathcal{G}%
_{i}^{j}=0$ and $\mathcal{G}_{i}^{i}-\mathcal{G}_{j}^{j}=0$ and therefore
one obtains the following constraints on the Weyl tensor:
\begin{equation}
0=\frac{2\hat{\alpha}_{2}{\tilde{C}_{ki}}^{nl}{\tilde{C}_{nl}}^{kj}}{r^{4}}-%
\frac{3\hat{\alpha}_{3}(2\tilde{C}^{nmkl}\tilde{C}_{klpm}{\tilde{C}^{pj}}%
_{ni}-4{\tilde{C}^{nm}}_{pk}{\tilde{C}^{kl}}_{ni}{\tilde{C}^{pj}}_{ml}+%
\tilde{C}^{jnkl}\tilde{C}_{klpm}{\tilde{C}^{pm}}_{ni})}{r^{6}},\text{ \ \ \ }%
(i\neq j)  \label{Gij0}
\end{equation}%
\begin{equation}
{\tilde{C}_{ki}}^{nl}{\tilde{C}_{nl}}^{kj}=\frac{1}{n}{\delta _{i}}^{j}{%
\tilde{C}_{km}}^{pq}{\tilde{C}_{pq}}^{km}\equiv \eta _{2}{\delta _{i}}^{j},
\label{theta}
\end{equation}%
\begin{eqnarray}
&&2(4{\tilde{C}^{nm}}_{pk}{\tilde{C}^{kl}}_{ni}{\tilde{C}^{pj}}_{ml}+{\tilde{%
C}^{pm}}_{in}\tilde{C}^{jnkl}\tilde{C}_{klpm})  \notag \\
&=&\frac{2}{n}{\delta _{i}}^{j}\left( 4{\tilde{C}^{qm}}_{pk}{\tilde{C}^{kl}}%
_{qr}{\tilde{C}^{pr}}_{ml}+{\tilde{C}^{pm}}_{qr}\tilde{C}^{qrkl}\tilde{C}%
_{klpm}\right)   \notag \\
&\equiv &\eta _{3}{\delta _{i}}^{j}.  \label{eta}
\end{eqnarray}%
Here, we pause to add some comments about the expected patterns of
conditions in $k$th-order Lovelock gravity. Comparing conditions (\ref%
{theta}) and (\ref{eta}) with the second- and third-order Lovelock
Lagrangians, respectively and using the expression of Lovelock Lagrangian
\cite{Lovelock}, one may expect that the conditions on
the Weyl tensor of base manifold are:%
\begin{equation}
\delta _{j_{1}j_{2}...j_{2p-1}j_{2p}}^{i_{1}i_{2}...i_{2p-1}i_{2p}}{\tilde{C}%
}_{i_{1}i_{2}}^{j_{1}j_{2}}...{\tilde{C}}_{i_{2p-1}i_{2p}}^{j_{2p-1j_{2p}}}%
\propto \eta _{p}, \text{\ \ \ } p=2...k.
\end{equation}%
Of course, one should note that $\tilde{C}_{jk}^{ik}=\tilde{C}%
_{ij}^{ij}=0$. For instance, the only term in the Gauss-Bonnet
Lagrangian which is nonzero is $\tilde{C}_{ki}^{nl}\tilde{C}_{nl}^{kj}$%
and the nonvanishing part of the third-order Lovelock is $8\tilde{C}_{pk}^{qm}\tilde{C}%
_{qr}^{kl}\tilde{C}_{ml}^{pr}+2\tilde{C}_{qr}^{pm}\tilde{C}^{qrkl}\tilde{C}%
_{klpm}$.

Getting use of these definitions, the $tt$ and $rr$ components of field
equation (\ref{Geq}) in vacuum reduce to
\begin{eqnarray}
0 &=&{\mathcal{G}_{t}}^{t}=\frac{(n-2)}{2r^{6}g^{4}}\{[r^{4}g^{2}+3\hat{%
\alpha}_{3}\hat{\eta}_{2}g^{2}+2\hat{\alpha}_{2}r^{2}(kg-1)g+3\hat{\alpha}%
_{3}(kg-1)^{2}]rg^{^{\prime }}+(kg-1)[(n-3)r^{4}g^{2}  \notag \\
&&+3(n-7)\hat{\alpha}_{3}\hat{\eta}_{2}g^{2}+(n-5)\hat{\alpha}%
_{2}r^{2}(kg-1)g+(n-7)\hat{\alpha}_{3}(kg-1)^{2}]g  \notag \\
&&+\left( (n-1)\hat{\alpha}_{0}+\frac{(n-5)\hat{\alpha}_{2}\hat{\eta}_{2}}{%
r^{4}}+\frac{(n-7)\hat{\alpha}_{3}\hat{\eta}_{3}}{r^{6}}\right) r^{6}g^{4}\},
\label{Gtt}
\end{eqnarray}%
\begin{eqnarray}
0 &=&{\mathcal{G}_{r}}^{r}=\frac{(n-2)}{2r^{6}fg^{3}}\{[r^{4}g^{2}+3\hat{%
\alpha}_{3}\hat{\eta}_{2}g^{2}+2\hat{\alpha}_{2}r^{2}(kg-1)g+3\hat{\alpha}%
_{3}(kg-1)^{2}]rf^{\prime }-(kg-1)[(n-3)r^{4}g^{2}  \notag \\
&&+3(n-7)\hat{\alpha}_{3}\hat{\eta}_{2}g^{2}+(n-5)\hat{\alpha}%
_{2}r^{2}(kg-1)g+(n-7)\hat{\alpha}_{3}(kg-1)^{2}]f  \notag \\
&&+\left( (n-1)\hat{\alpha}_{0}+\frac{\hat{\alpha}_{2}(n-5)\hat{\eta}_{2}}{%
r^{4}}+\frac{(n-7)\hat{\alpha}_{3}\hat{\eta}}{r^{6}}\right) r^{6}g^{4}\},
\label{Grr}
\end{eqnarray}%
where we have used the definition $\hat{\eta}_{2}=(n-6)!\eta _{2}/(n-2)!$
and $\hat{\eta}_{3}=(n-8)!\eta _{3}/(n-2)!$ for simplicity. It is notable to
mention that for these kinds of Einstein metrics $\hat{\eta}_{2}$ is always
positive, but $\hat{\eta}_{3}$ can be positive or negative relating to the
metric of the spacetime. As an example, the manifolds that are
cross-products of $p$ $(p\geq 3)$ of two-hyperbola ($\mathcal{H}^{2}$) are
Einstein manifolds with negative $\hat{\eta}_{3}$. The vacuum equation $%
\mathcal{G}_{t}^{t}-\mathcal{G}_{r}^{r}=0$ implies that $d(fg)/dr=0,$ and
therefore one can take $g(r)=1/f(r)$ by rescaling the time coordinate $t$.
Introducing

\begin{equation}
\psi (r)=\frac{\kappa -f(r)}{r^{2}},  \label{Psi}
\end{equation}%
we find that the remaining equations admit a solution if $\hat{\eta}_{2}$
and $\hat{\eta}_3$ defined in Eqs. (\ref{theta}) and (\ref{eta}) are
constant and $\psi (r)$ satisfies
\begin{equation*}
\left\{ r^{n-1}\left[ \hat{\alpha}_{3}\psi ^{3}+\hat{\alpha}_{2}\psi
^{2}+\left( 1+\frac{3\hat{\alpha}_{3}\hat{\eta}_{2}}{r^{4}}\right) \psi +%
\hat{\alpha}_{0}+\frac{\hat{\alpha}_{2}\hat{\eta}_{2}}{r^{4}}+\frac{\hat{%
\alpha}_{3}\hat{\eta}_3}{r^{6}}\right] \right\}^{\prime}=0.
\end{equation*}%
Integrating the above equation, one obtains

\begin{equation}
\left( 1+\frac{3\hat{\alpha}_{3}\hat{\eta}_{2}}{r^{4}}\right) \psi +\hat{%
\alpha}_{2}\psi ^{2}+\hat{\alpha}_{3}\psi ^{3}+\hat{\alpha}_{0}+\frac{\hat{%
\alpha}_{2}\hat{\eta}_{2}}{r^{4}}+\frac{\hat{\alpha}_{3}\hat{\eta}_3}{r^{6}}-%
\frac{m}{r^{n-1}}=0,  \label{Eq3}
\end{equation}%
where $m$ is the integration constant known as the mass parameter. One may
note that Eq. (\ref{Eq3}) reduces to the algebraic equation of Lovelock
gravity for constant-curvature horizon when $\hat{\eta}_{2}=\hat{\eta}_3=0$.

The mass density, the mass per unit volume $V_{n-2}$, associated to the
spacetime may be written as
\begin{equation}
M=\frac{(n-2)r_{h}^{n-1}}{16\pi }\left\{ \hat{\alpha}_{0}+\frac{\kappa }{%
r_{h}^{2}}+\frac{\hat{\alpha}_{2}}{r_{h}^{4}}(\kappa ^{2}+\hat{\eta}_{2})+%
\frac{\hat{\alpha}_{3}}{r_{h}^{6}}(\kappa ^{3}+3\kappa \hat{\eta}_{2}+\hat{%
\eta}_{3})\right\} .  \label{massP}
\end{equation}

\section{Black Hole Solutions \label{Bla}}

One may note that in order to have the effects of nonconstancy of the
curvature of the horizon in third-order Lovelock gravity, $n$ should be
larger than $7$. This can be seen in the definition of $\hat{\eta}_{3}$,
which is zero for $n\leq 7$. A general solution of this equation can be
written as
\begin{eqnarray}
f(r) &=&\kappa +\frac{\hat{\alpha}_{2}r^{2}}{3\hat{\alpha}_{3}}\left\{
1+\left( j(r)\pm \sqrt{h+j^{2}(r)}\right) ^{1/3}-h^{1/3}\left( j(r)\pm \sqrt{%
h+j^{2}(r)}\right) ^{-1/3}\right\} ,  \notag \\
j(r) &=&1-\frac{9\hat{\alpha}_{3}}{2\hat{\alpha}_{2}^{2}}+\frac{27\hat{\alpha%
}_{3}^{2}}{2\hat{\alpha}_{2}^{3}}\left( \hat{\alpha}_{0}-\frac{m}{r^{n-1}}%
+\frac{\hat{\alpha}_{3}\hat{\eta}_{3}}{r^{6}}\right) ,\text{ \ }  \notag \\
\text{\ \ \ \ }h &=&\left( -1+\frac{3\hat{\alpha}_{3}}{\hat{\alpha}_{2}^{2}}+%
\frac{9\hat{\alpha}_{3}^{2}\hat{\eta}_{2}}{\hat{\alpha}_{2}^{2}r^{4}}\right)
^{3},  \label{fstat}
\end{eqnarray}%
These are the most general solutions of the third-order Lovelock equation in
vacuum with the conditions (\ref{theta}) and (\ref{eta}) on their boundaries
$\mathcal{K}^{(n-2)}$.

First, we investigate the asymptotic behavior of this solution. The
asymptotic behavior of the solutions is the same as those with
constant-curvature horizons. This is due to the fact that Eq. (\ref{Eq3}) at
very large $r$\ reduces to
\begin{equation}
\hat{\alpha}_{3}\psi _{\infty }^{3}+\hat{\alpha}_{2}\psi _{\infty }^{2}+\psi
_{\infty }+\hat{\alpha}_{0}=0,  \label{Asym}
\end{equation}%
which is exactly the same as third-order Lovelock or quasitopological cubic
gravity \cite{Myers}. One may note that in the absence of the cosmological
constant ($\hat{\alpha}_{0}=0$), the solution is asymptotically flat provided $%
\kappa =1$. This can be noted by considering Eq. (\ref{Asym}) which has a
zero root for $\hat{\alpha}_{0}=0$. For $\hat{\alpha}_{0}=1$, the solution
is asymptotically AdS if Eq. (\ref{Asym}) has positive real roots. For more
details on the asymptotic behavior see \cite{Myers}. As in the case of black
hole solutions wdith constant-curvature horizon, the Kretschmann scalar $%
R_{\mu \nu \rho \sigma }R^{\mu \nu \rho \sigma }$ diverges at $r=0$. Since
the dominant term as $r$ goes to zero is $m/r^{n-1}$ for $n>7$, as in the
case of third-order Lovelock gravity with constant-curvature horizon, there
is an essential singularity located at $r=0$ which is spacelike. Note that
the radius of horizon is given by the largest real root of
\begin{equation}
\hat{\alpha}_{0}r_{h}^{n-1}+\kappa r_{h}^{n-3}+\hat{\alpha}_{2}(\kappa ^{2}+%
\hat{\eta}_{2})r_{h}^{n-5}+\hat{\alpha}_{3}(\kappa ^{zz3}+3\kappa \hat{\eta}%
_{2}+\hat{\eta}_{3})r_{h}^{n-7}-m=0,  \label{horizon}
\end{equation}%
where $r_{h}$ is the radius of horizon.

Here, we pause to give a few comments on the differences of the solutions of
third-order Lovelock gravity with constant and nonconstant-curvature
horizons. While the solutions of third order Lovelock gravity with constant
curvature horizon and $m=0$ is the AdS metric with no horizon, the solutions
of Lovelock gravity with nonconstant-curvature horizon and $m=0$\ are only
asymptotically AdS and may have horizon. This is due to the fact that $\hat{%
\eta}_{3}$ can be negative and therefore Eq. (\ref{horizon}) can have a real
positive root. Moreover, in third-order Lovelock gravity with constant-
curvature horizon $h$ can be zero for $\hat{\alpha}_{3}=\hat{\alpha}%
_{2}^{2}/3$ and therefore the solution may be written in the simpler form:%
\begin{eqnarray}
f(r) &=&\kappa +\frac{r^{2}}{\hat{\alpha}_{2}}\left\{
1-[2j(r)]^{1/3}\right\} ,  \notag \\
j(r) &=&-\frac{1}{2}+\frac{3\hat{\alpha}_{2}}{2}\left( \hat{\alpha}_{0}-%
\frac{m}{r^{n-1}}\right) ,  \label{Spe}
\end{eqnarray}%
while for the case of nonconstant curvature, $h$ cannot be zero and
therefore we cannot have this special kind of solution.

\section{Thermodynamics of The black hole solutions \label{The}}

Using the relation between the temperature and surface gravity, the Hawking
temperature of the black hole is obtained to be
\begin{equation}
T=\frac{f^{\prime }(r_{h})}{4\pi }=\frac{(n-1)r_{h}^{6}\hat{\alpha}%
_{0}+(n-3)\kappa r_{h}^{4}+(n-5)\hat{\alpha}_{2}(\hat{\eta}_{2}+\kappa
^{2})r_{h}^{2}+(n-7)\hat{\alpha}_{3}(\hat{\eta}_{3}+3\kappa \hat{\eta}%
_{2}+\kappa ^{3})}{4\pi r_{h}[r_{h}^{4}+2\kappa \hat{\alpha}_{2}r_{h}^{2}+3%
\hat{\alpha}_{3}(\hat{\eta}_{2}+\kappa ^{2})]}.  \label{Temp}
\end{equation}%
Due to the fact that $\hat{%
\eta}_{3}$ can be negative, it is apparent from Eq. (\ref{Temp}) that a
degenerate Killing horizon can exist for $\kappa \geq 0$ and therefore one
may have an extreme black hole. This feature does not happen for the solutions
of third-order Lovelock gravity with constant-curvature horizons \cite%
{Dehghani} or second-order Lovelock gravity with constant or nonconstant-
curvature horizons \cite{Maeda}.

In higher curvature gravity the area law of entropy, which states that the
black hole entropy equals one-quarter of the horizon area \cite{Beckennstein}%
, is not satisfied \cite{Lu}. One approach to calculate the entropy is
through the use of the Wald prescription which is applicable for any black hole
solution whose event horizon is a Killing one \cite{Wald}. The Wald entropy
may be written as
\begin{equation}
S=-2\pi \oint d^{n-2}x\sqrt{\gamma }\sum_{p=1}^{3}Y_{p},\text{ \ \ \ \ \ }%
Y_{p}=Y_{p}^{\mu \nu \rho \sigma }\hat{\varepsilon}_{\mu \nu }\hat{%
\varepsilon}_{\rho \sigma },\text{\ \ \ \ \ \ }Y_{p}^{\mu \nu \rho \sigma }=%
\frac{\partial \mathcal{L}^{(p)}}{\partial R_{\mu \nu \rho \sigma }},
\label{entropy}
\end{equation}%
where $\hat{\varepsilon}_{\mu \nu }$ is the binormal to the horizon and $%
\mathcal{L}^{(p)}$ is the $p$th-order Lovelock Lagrangian. Following the
given description, $Y_{1}$ and $Y_{2}$ are \cite{Myers}
\begin{equation}
Y_{1}=-\frac{1}{8\pi }  \label{Ein-entropy}
\end{equation}%
\begin{equation}
Y_{2}=-\frac{\hat{\alpha}_{2}}{4\pi }[R-2(R_{t}^{t}+R_{r}^{r})+2R_{tr}^{tr}]
\label{Gauss-entropy}
\end{equation}%
Also we calculate $Y_{3}$ to be
\begin{eqnarray}
Y_{3} &=&-\frac{3\hat{\alpha}_{3}}{4\pi }\{-12(R{^{tm}}_{tn}R{^{rn}}_{rm}-R{%
^{tm}}_{rn}R{{^{r}}_{mt}}^{n})+12R^{trmn}R_{trmn}-24[R{^{tr}}_{tm}R{_{r}}%
^{m}-R{^{tr}}_{rm}{R_{t}}^{m}  \notag \\
&&+\frac{1}{4}(R_{mnpr}R^{mnpr}+R_{mnpt}R^{mnpt})]+3(2R{R^{tr}}_{tr}+\frac{1%
}{2}R_{mnpq}R^{mnpq})  \notag \\
&&+12({R^{t}}_{t}{R^{r}}_{r}-{R^{t}}_{r}{R^{r}}_{t}+{R^{r}}_{mrn}R^{mn}+{%
R^{t}}_{mtn}R^{mn})+12(R^{rm}R_{rm}+R^{tm}R_{tm})  \notag \\
&&-6[R_{mn}R^{mn}+R({R^{r}}_{r}+{R^{t}}_{t})]+\frac{3}{2}R^{2}\}.
\label{Lov-entropy}
\end{eqnarray}%
Using Eq. (\ref{Riem Ten}), one can calculate the entropy density for
nonconstant-curvature manifold in third-order Lovelock gravity to be
\begin{equation}
S=-2\pi \{Y_{1}+Y_{2}+Y_{3}\}=\frac{r_{h}^{n-2}}{4}\left\{ 1+\frac{2\kappa
\hat{\alpha}_{2}(n-2)}{r_{h}^{2}(n-4)}+\frac{3\hat{\alpha}_{3}(n-2)(\hat{\eta%
}_{2}+\kappa ^{2})}{r_{h}^{4}(n-6)}\right\} .  \label{Entro}
\end{equation}%
We see that $\hat{\eta}_{2}$ appears in the entropy and therefore the the
nonconstancy of the horizon affects the entropy of the black hole. This
does not happen for the Gauss-Bonnet solution. One should notice that the
entropy calculated in Eq. (\ref{Entro}) could also be obtained using the
relation
\begin{equation}
S=\frac{1}{4}\sum\limits_{q=1}^{p}p\hat{\alpha}_{p}\int d^{n-2}x\sqrt{\gamma
}\mathcal{\tilde{L}}^{(p-1)}
\end{equation}%
introduced in \cite{Jacobson}, where $\gamma $ is the determinant of induced
metric and $\mathcal{\tilde{L}}^{(p)}$ is the $p$th-order Lovelock
Lagrangian of the metric $\gamma _{ij}$. Getting use of Eqs. (\ref{massP}%
), (\ref{Temp}) and (\ref{Entro}), we obtain $\partial {M}=T\partial {S}$
and therefore the first law of black hole thermodynamics is satisfied.

\section{Stability of Black holes with $\protect\kappa =0$ \label{Stab}}

\begin{figure}[tbp]
\centering {\includegraphics[width=7cm]{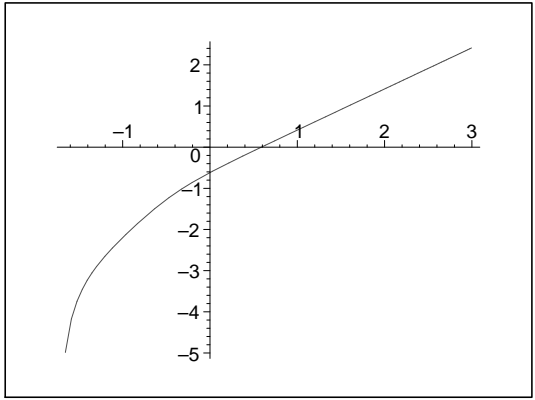}}
\caption{The temperature versus horizon radius in logarithmic scale for $n=8$%
, $\hat{\protect\alpha}_{0}=1$, $\hat{\protect\alpha}_{2}=0.2$, $\hat{%
\protect\alpha}_{3}=0.05$, $\hat{\protect\eta}_2=0.5$ and $\hat{\protect\eta}%
_3=-0.2$.}
\label{extreme}
\end{figure}

It is known that the black holes of Lovelock gravity with zero curvature
horizon are stable \cite{Dehghani}. Here, we investigate the stability of
black holes of Lovelock gravity with nonconstant-curvature horizons and
give some special features of $\kappa =0$ black hole solutions with
non-constant curvature horizon, which are drastically different from $\kappa
=0$ solutions with constant-curvature horizon. In the case of $\kappa =0$,
the entropy density of the black hole reduces to
\begin{equation}
S_{0}=\frac{r_{h}^{n-2}}{4}\left\{ 1+\frac{3\hat{\alpha}_{3}(n-2)\hat{\eta}%
_{2}}{r_{h}^{4}(n-6)}\right\} .  \label{Sn0}
\end{equation}%
Since $\hat{\eta}_{2}\neq 0$, the black holes with nonconstant-curvature
horizon do not obey the area law of entropy, while the entropy of $\kappa =0$
\ black holes with constant-curvature horizon obey the area law. The
temperature of such a black hole is
\begin{equation}
T_{0}=\frac{(n-1)r_{h}^{6}\hat{\alpha}_{0}+(n-5)\hat{\alpha}_{2}\hat{\eta}%
_{2}r_{h}^{2}+(n-7)\hat{\alpha}_{3}\hat{\eta}_{3}}{4\pi r_{h}(r_{h}^{4}+3%
\hat{\alpha}_{3}\hat{\eta}_{2})},  \label{Tn0}
\end{equation}%
where $\hat{\alpha}_{0}=1$ and $0$ for asymptotically AdS and flat
solutions, respectively. It is worth noting that for $\hat{\eta}_3 =\hat{\eta%
}_{3ext}$:
\begin{equation*}
\hat{\eta}_{3ext}=-\frac{(n-1)r_{h}^{6}\hat{\alpha}_{0}+(n-5)\hat{\alpha}_{2}%
\hat{\eta}_{2}r_{h}^{2}}{(n-7)\hat{\alpha}_{3}}
\end{equation*}%
the temperature can be zero, and therefore in contrast to the case of $%
\kappa =0$ of Lovelock black holes, extreme black holes may exist. This can
be seen in Fig. \ref{extreme}.

\begin{figure}[tbp]
\centering {\includegraphics[width=7cm]{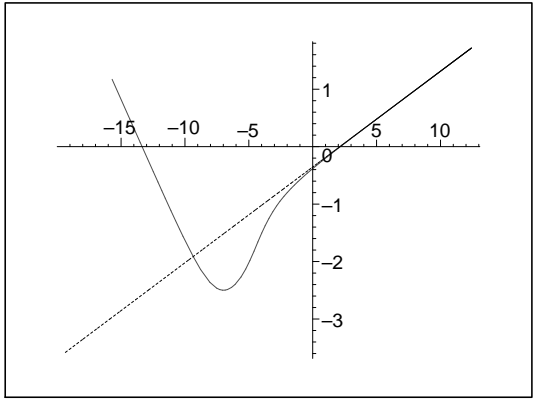}}
\caption{The temperature versus entropy in logarithmic scale for $n=8$%
, $\hat{\protect\alpha}_{0}=1$, $\hat{\protect\alpha}_{2}=0.2$ and $\hat{%
\protect\alpha}_{3}=0.05$ with $\hat{\protect\eta}_2=0.5$ and $\hat{\protect%
\eta}_3=0.1$ (line), and $\hat{\protect\eta}_2=\hat{\protect\eta}_3=0$
(dotted).}
\label{Lstab}
\end{figure}

\begin{figure}[tbp]
\centering {\includegraphics[width=7cm]{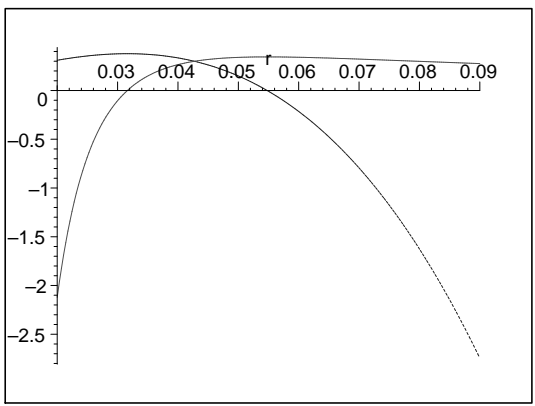}}
\caption{$10^{-2}dT/dS$ (line) and $10^6F$ (dotted) versus $r_{h}$ for $n=8$, $%
\hat{\protect\alpha}_{0}=1$, $\hat{\protect\alpha}_{2}=0.2$, $\hat{\protect%
\alpha}_{3}=0.05$, $\hat{\protect\eta}_2=0.5$ and $\hat{\protect\eta}_3=0.006
$. }
\label{Gstab}
\end{figure}

The local stability of a thermodynamic system may be performed by analyzing
the curve of $T$ versus $S$. Figure \ref{Lstab} depicts $\log T$ versus $%
\log S$ and shows that small black holes are unstable for positive $\hat{\eta}_3>%
\hat{\eta}_{3ext}$, while the very large black holes with nonconstant-
curvature horizon are the same as black holes with constant curvature.
To analyze the global stability, we should check the free energy of the
black hole which is defined by $F\equiv M-TS$, whereby negative value
ensures global stability \cite{Hawking}. Substituting the expressions for
mass, temperature and entropy from Eqs. (\ref{massP}), (\ref{Tn0}) and (\ref%
{Sn0}), one can perform the analysis of global stability. We plot the free
energy versus the radius of black holes for $\hat{\eta}_3>\hat{\eta}_{3ext}$
in Fig. \ref{Gstab} which shows that small black holes are unstable both
locally and globally, while there are medium black holes which may be
locally stable, but they are globally unstable. So, in contrast to the case
of $\kappa =0$ black holes with constant-curvature horizon which are stable
\cite{Dehghani}, the black hole solutions here may have unstable phases both
locally and globally.

\section{Concluding Remarks}

In this paper, we assumed that the $n$-dimensional spacetime is a cross
product of the two-dimensional Lorentzian spacetime and an $(n-2)$%
-dimensional nonconstant space. We found that the nontrivial Weyl tensor
of such exotic horizons is exposed to the bulk dynamics through the higher-
order Lovelock terms, severely constraining the allowed horizon geometries
and adding a novel chargelike parameter to the black hole potential. Indeed,
we found that the third-order Lovelock gravity can have a new class of black
hole solutions with nonconstant-curvature horizons provided one imposes two
conditions on the Weyl tensor. The first condition is the one which has been
introduced by Dotti and Gleiser \cite{Dotti} in Gauss-Bonnet gravity, while
the second one is an additional condition involving the Weyl tensor of the
horizon manifold with the advantage of higher curvature terms appearing in
third-order Lovelock equations. This leads to a new class of static
asymptotically flat and (A)dS black hole solutions. It is worth
comparing our result with the already existing results in the literature.
First, while in EGB gravity with an arbitrary Gauss-Bonnet coefficient only one
condition is imposed on the Weyl tensor of Einstein horizon \cite{Dotti},
here we faced with two conditions on the Weyl tensor of Einstein horizon.
Second, as in the case of Gauss-Bonnet gravity with an arbitrary coupling
constant we found that the horizon should be an Einstein manifold. However, for
the cases when there is a unique maximally symmetric solution, the base
manifold acquires more freedom \cite{Dot2,Bog,Oliv1,Can,Oliv2}. Third, while
only one parameter appears in the solutions of Lovelock gravity in the
Chern-Simons case \cite{Oliv1} or third- and higher-order Lovelock
Born-Infeld gravity \cite{Can, Oliv2}, here we encountered with two new
different chargelike parameters $\hat{\eta}_{2}$ and $\hat{\eta}_{3}$.
Thus, one may expect that in the case of Lovelock gravity with arbitrary
Lovelock coefficients the number of charge-like parameters $\hat{\eta}_{p}$%
's will increase as the order of Lovelock gravity becomes larger.

The thermodynamics of these black hole solutions have been investigated by
calculating the temperature and the entropy through the use of the Wald formula.
We found that the thermodynamic quantities satisfy the first law of
thermodynamics. In contrast to the black holes of Lovelock gravity with
constant curvature horizons, we found that there may exist extreme black
holes with nonconstant-curvature horizon. We also found that the effect of
the Weyl tensor in the metric and the expressions for temperature and
entropy, lead to new features of the solutions that do not appear for the
solutions of second-order Lovelock gravity with nonconstant-curvature
horizons or in higher-order Lovelock gravity with constant-curvature ones.
In order to show the main differences of black holes with constant- and
nonconstant-curvature horizons, we went through the details of
thermodynamics and stability analysis of black holes with $\kappa =0$.
First, we found that the entropy does not obey the area law as a consequence
of $\hat{\eta}_{2}$ appearing in the entropy expression. Second, we found
that there may exist extreme black hole. Moreover, black holes with
nonconstant-curvature horizons have an unstable phase both locally and
globally, while those with constant-curvature horizons are stable both
locally and globally \cite{Dehghani}. But, it is worth mentioning that very
large black holes with constant- and nonconstant-curvature horizons have the
same features.

\acknowledgments{N. Farhangkhah thanks the Research Council of
Shiraz Branch of Islamic Azad University. This work has been supported
financially by Research Institute for Astronomy \& Astrophysics of Maragha
(RIAAM), Iran.}

\end{document}